\documentclass[aps,twocolumn,pra,showpacs,floatfix]{revtex4}
\usepackage{epsfig}
\usepackage{dcolumn}

\begin{document}

\title{\bf Actinide ions for testing the spatial $\alpha-$variation hypothesis}

\author{V. A. Dzuba$^1$, M. S. Safronova$^{2,3}$, U. I. Safronova$^{4}$,
  and V. V. Flambaum$^1$}
\affiliation{$^1$School of Physics, University of New South Wales,
Sydney 2052, Australia}
\affiliation{$^2$Department of Physics and Astronomy, University of Delaware, Newark, Delaware 19716, USA}
\affiliation{$^3$Joint Quantum Institute, National Institute of Standards and Technology and the University of Maryland, College Park, 20742, USA}
\affiliation{$^4$University of Nevada, Reno, Nevada 89557, USA}

\date{\today}
\begin{abstract}
Testing the spatial variation of fine-structure constant $\alpha$
indicated in [Webb et al., Phys. Rev. Lett. 107, 191101 (2011)] with
terrestrial laboratory atomic  measurements  requires at least
$\dot{\alpha}/\alpha \sim 10^{-19}~\textrm{y}^{-1}$ sensitivity. 
 We conduct a systematic search of atomic systems for such a test that
 have all features of the best optical clock transitions leading to
 possibility of the frequency measurements with fractional accuracy on
 the level of $10^{-18}$ or better and have a factor of 100 extra
 enhancement of $\alpha$-variation in comparisons to experimental
 frequency ratio measurement accuracy. We identify the pair of
 actinide Cf$^{15+}$ and Es$^{16+}$ 
 ions as the best system for a test of spatial $\alpha-$variation
 hypothesis as it satisfies both of these requirements and have
 sufficiently simple electronic structure to allow for high-precision
 predictions of all atomic properties required for rapid experimental
 progress. 
 \end{abstract}
\maketitle

Theories aimed at unifying gravity with other fundamental
interactions~\cite{Uzan,Uzan11} suggest variation 
of the fundamental constants, such as fine-structure constant
$\alpha=e^2/\hbar c$ that characterises the strength of the 
electromagnetic interaction.
In 2011, a very large analysis of the quasar absorption spectra that combined
data taken by Keck telescope in Hawaii and the Very Large Telescope
(VLT) in Chile indicated $4\sigma$ spatial 
gradient in the value of $\alpha$ \cite{alpha-dipole}. Such result,
implying that $\alpha$ was larger in the past in one 
direction, but smaller in the opposite direction, would have a
profound significance for the ``fine-tuning'' problem: the values of
fundamental constants that can support life happen to fall within a
narrow range \cite{dipole-dot}. 
2015 study of systematic distortions in the wavelengths scales of
high-resolution spectrographs \cite{WhitMur15} showed that 
instrumental error may weaken the spatial variation result
\cite{alpha-dipole}, but can not explain all of the observed 
variation of $\alpha$. Therefore, it became paramount to set up a
laboratory experiment 
capable of potential observation of the spatial $\alpha$-variation at
the level indicated by \cite{alpha-dipole}. 

Frequencies of atomic transitions depend on $\alpha$, therefore
long-term monitoring of 
frequency ratio with ultra-high precision atomic clocks enables tests
of $\alpha$-variation at the present time. 
We also note a special case of non-clock test of $\alpha$-variation
in Dy \cite{Dy-alpha-dot}, which uses extremely high sensitivity of a
specific transition between two almost degenerate excited states while
accuracy of the frequency measurements is significantly lower than in
atomic clocks.  
Current laboratory limit for time variation of the fine
structure constant stands on the level of $2\times 10^{-17}$
fractional change per 
year~\cite{Al+,Dy-alpha-dot,Gill-Yb+14,Peik-Yb+14}. However, test the
spatial alpha-dipole 
hypothesis presented in \cite{alpha-dipole} requires at least two
orders of magnitude improvement in accuracy 
of laboratory measurements \cite{dipole-dot}, $\dot{\alpha}/\alpha
\sim 10^{-19}~\textrm{y}^{-1}$  due to a relatively small velocity of
Sun 
relative to the cosmic microwave background, 369~km/s.
 Moreover, unambiguous separation of the time-variation and
 space-variation of $\alpha$ would require observation of the annual
 modulation of the signal with Earth motion around the Sun, which may
 become possible only with   $\dot{\alpha}/\alpha \sim
 10^{-20}~\textrm{y}^{-1}$ 
sensitivity \cite{dipole-dot}. Review of present limits of spatial
variation of fundamental constants, which discusses both cosmological
scale and local scales regimes is given in \cite{Uzan11}. 

The worlds best optical lattice atomic clocks
  approach fractional accuracy of
10$^{-18}$~\cite{NicCamHut15,Sr-clock14,Katori}, with the smallest
uncertainty of $2\times10^{-18}$ achieved for Sr
\cite{NicCamHut15}. However, this does not immediately translates 
into high sensitivity to $\alpha$-variation, because Sr and Yb clock
atomic transitions are not sufficiently sensitive to the variation of
the fine structure constant~\cite{AngDzuFla04}. 

 The dependence of atomic frequencies on $\alpha$ can be parameterized
 by the sensitivity coefficient $q$~\cite{DzuFlaWeb99,DzuFlaWeb99a} 
\begin{equation}
\omega(x) = \omega_0 +qx,
\label{eq:wx}
\end{equation}
which can be rather accurately  determined from theoretical computations.
Here,
$$ x=\left[\left(\frac{\alpha}{\alpha_0}\right)^2-1\right]$$
and $\alpha_0$ is a current value of $\alpha$.
The parameter $q$ links variation of frequency $\omega$ to the
variation of $\alpha$ 
\begin{equation}
\frac{\delta \omega}{\omega_0} = \frac{2q}{\omega_0}\frac{\delta
  \alpha}{\alpha_0} \equiv K  \frac{\delta \alpha}{\alpha_0}, 
\label{eq:K}
\end{equation}
where $K=2q/\omega_0$ is an enhancement factor.
In atomic clock laboratory $\alpha$-variation tests, the ratio of two
clock frequencies is monitored, and the 
sensitivity of such test  is then described by the difference in their
respective $K$ values, i.e. $\Delta K=|K_2-K_1|$: 
\begin{equation}
\frac{\partial}{\partial t}\ln\frac{\omega_2}{\omega_1} = \left(K_2 -
  K_1\right)\frac{1}{\alpha}\frac{\partial\alpha}{\partial t}
\label{e:ir17}
\end{equation}

  The $K$ factor for Sr clock transitions is 0.06~\cite{FlaDzuCJP}. In
  fact, the $K$ factors for most 
 of the clocks currently in development, Mg, Al$^+$, Ca$^+$, Sr$^+$,
 Sr, Yb, Yb$^+$ quadrupole transition, and Hg are very small,
 $0.008-1.0$~\cite{FlaDzuCJP}. 
  The $K$ factors for Hg$^+$ clock and Yb$^+$ octupole clock transitions
are $-2.9$ and $-6$, making them the best candidates for one part of
the clock comparison pair against the clock from the 
previous group that will essentially serve as a reference with little
sensitivity to $\alpha$-variation. 
As a result, reaching  $\dot{\alpha}/\alpha \sim
10^{-20}~\textrm{y}^{-1}$ precision with any of these clock 
combinations with require better than 10$^{-19}$ precision of the
clock ratio, even with one of the 
clocks based on octupole Yb$^+$ transitions.
Comparison of different clocks will become even more challenging
beyond 10$^{-18}$ accuracy due to sensitivity to the environment,
including temperature and gravitational 
potential ~\cite{LudBoyYe15}. For example, a clock on the surface of
the Earth that is higher by just 1~cm than 
another identical clock runs faster by $\delta \omega/\omega_0\sim
10^{-18}$ ~\cite{LudBoyYe15}.

Therefore, it is highly desirable to find a combination of two
transitions which  can be used to design of a high-precision clocks
and has at least $\Delta K=100$, which is a subject of this work.
The transitions which combine high sensitivity to $\alpha$-variation
with potential to design very accurate optical clock   can be found in
highly charged ions (HCI) 
~\cite{HCI-PRL14,BDFO-Cf12,BDF-HCI10,Ho14+15} or in a unique case of
$^{229}$Th nucleus 
~\cite{Fla-Th229-05}.
The use of the HCI for search of the time
variation of the fine structure constant $\alpha$ ($\alpha = e^2/\hbar
c$) was considered in 
Refs.~\cite{BDF-HCI10,BDFO-hole11,BDFO-Cf12,DDF12,HCI-PRL14,Ag-likeHCI14,Cd-likeHCI14,IrPRL15}.
Sympathetic cooling of HCI ions with Be$^+$ has just been demonstrated in \cite{SchVerSch15}, paving the way to the future clock development.

We have studied a large variety of highly-charged ions systems  to identify the systems
with $\Delta K >100$ that approximately satisfy
the criteria for good clock transitions  formulated in Ref.~\cite{Ho14+15}:
\begin{itemize}
\item The transition  is in optical region ($230 \ \rm{nm} < \lambda < 2000 \ \rm{nm}$
  or $5000 \ \rm{cm}^{-1} < \hbar \omega < 43000 \ \rm{cm}^{-1}$).
\item The lifetime of the clock state is  between 100 and $\sim10^4$ seconds.
\item There are other relatively strong optical transitions
  (equivalent lifetime on the order of  $\tau \lesssim 1$ ms).
\item Clock transition is not sensitive to perturbations caused by the black-body radiation (BBR), gradients of external electric fields, etc.
\end{itemize}

\begin{figure}
\epsfig{figure=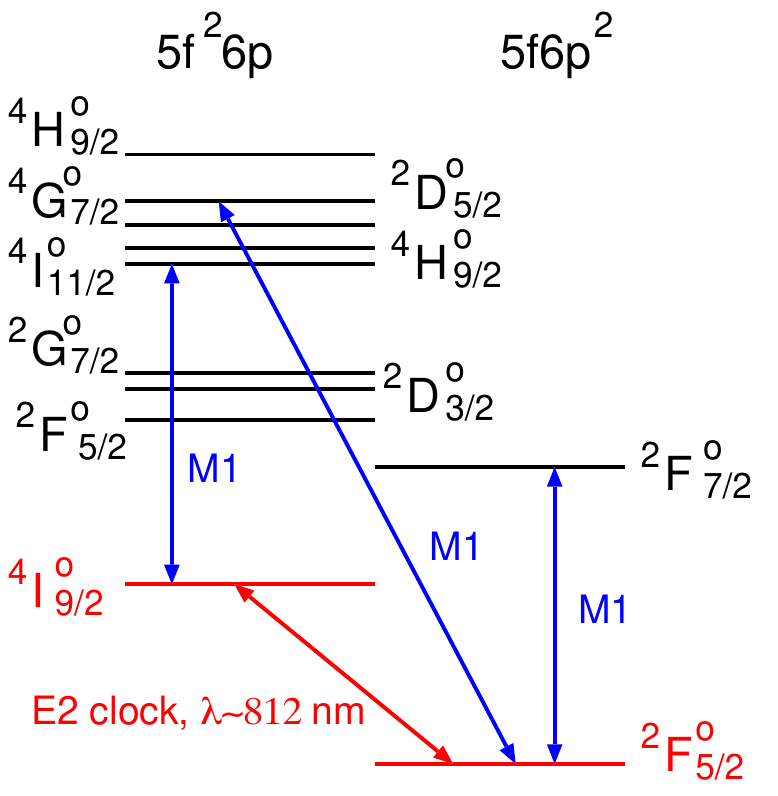,scale=0.9}
\caption{Low-lying energy levels of Cf$^{15+}$ (roughly in
  scale). Leading configurations are shown on the top. Clock
  transition is the electric quadrupole (E2) transition between the ground
and first excited state. Strongest magnetic dipole (M1) transitions
are also shown.}
\label{f:Cf15}
\end{figure}
The largest sensitivity of atomic optical transitions to the variation of the fine structure constant has been
found so far in Cf$^{16+}$ ions~\cite{BDFO-Cf12}. However, the states of Cf$^{16+}$ sensitive to
variation of $\alpha$ lack important features of clock states listed above. First four excited states of Cf$^{16+}$ ion are metastable, however, the lifetimes of all these clock states are outside of the desirable value range, which will makes accurate measurements
problematic. In the present work, we find that the pair of Cf$^{15+}$ and  Es$^{16+}$ ions satisfy the clock requirements
with the enhanced sensitivity to $\alpha$-variation of $\Delta K=110$.
 Both of these ions
has first excited metastable state which has all features  of the upper clock state, such as convenient values of the
frequencies and transition rates, low sensitivity to external perturbations, etc. The transition between ground
and these metastable states corresponds to the $f-p$ single-electron transition. This makes the transitions
sensitive to the variation of the fine-structure constant.  An additional advantage comes from the fact that
there is a $5f-6p$ level crossing between Cf
 $^{15+}$ and Es$^{16+}$ ions, i.e. the ground state of Cf$^{15+}$ becomes upper clock state in $\textrm{Es}^{16+}$ and vice
 versa. This means that if $\alpha$ changes in time, the drift of clock frequencies in Cf$^{15+}$ and Es
 $^{16+}$ have different signs, e.g. if $\alpha$ increases, the clock frequency of Cf$^{15+}$ increases as well
 while clock frequency of Es$^{16+}$  decreases. This leads to extra enhancement of the sensitivity of the ratio
 of clock frequencies to the variation of the fine structure constant.

Another advantage of using Cf$^{15+}$ and Es$^{16+}$  ions is due to the fact that they represent
good compromise between simple electron structure and the abundance of relatively strong optical transitions
which can be used for cooling, detection, etc. Simple electron structure allows for reliable theoretical predictions
of the ion properties, which will tremendously simplify the spectra identification and experimental search for the clock transitions.
 Most of the systems
considered before have either simple electron structure~\cite{BDF-HCI10,BDFO-Cf12,DDF12,HCI-PRL14,
Ag-likeHCI14,Cd-likeHCI14} but not many strong optical transitions, or complicated electron structure (many
electrons in open shells)~\cite{BDFO-hole11,IrPRL15,Ho14+15} that leads to experimental difficulties.

\begin{figure}
\epsfig{figure=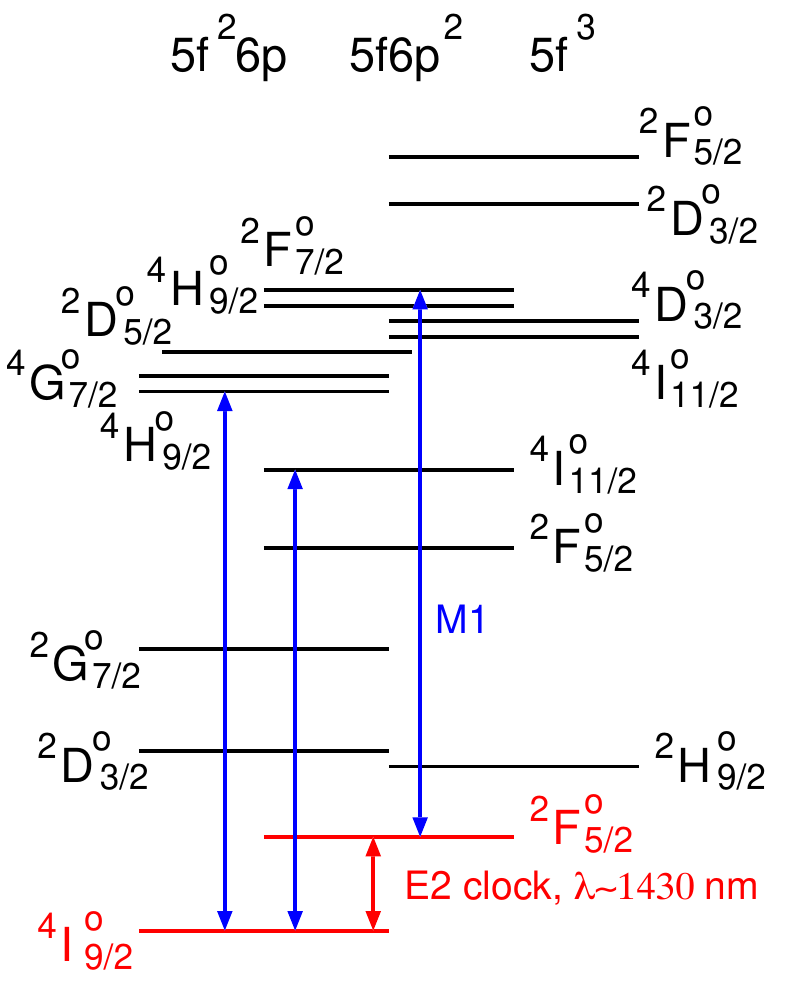,scale=0.9}
\caption{Low-lying energy levels of Es$^{16+}$, see caption for
  Fig.~\ref{f:Cf15}.}
\label{f:Es16}
\end{figure}

Neither californium nor einsteinium have stable isotopes. However,
they have very long-lived isotopes, such as $^{249}$Cf with half life
time of 351 years and $^{252}$Es with half life time of 1.3 years. There are many facilities around the globe
(Berkeley, Dubna, Darmstadt, RIKEN, etc.) which produce and study unstable isotopes (see,
e.g.~\cite{Cf-prod,Es-prod}).

Below, we described our predictions for the properties of these ions
relevant to the search for $\alpha$-variation. 
We use the CI+all-order method \cite{SD+CI,Dzu-CI-SD14}, which
combines the single-double linearised  coupled-cluster ~\cite{CC1} and
the 
configuration interaction approaches. The B-spline
technique~\cite{B-splines} is used to construct a set of
single-electron basis states in the $V^{N-3}$ 
approximation~\cite{Dzu05}, which means that the initial
self-consistent Hartree-Fock procedure is done for the 
Hg-like closed-shell core with three valence electrons
removed. Core-valence and core-core correlations are 
included with the use of the single-double coupled cluster method
while valence-valence correlations between 
the three electrons are
included with the use of the configuration interaction method. Breit
and quantum electrodynamic (QED) 
corrections are also included as described in Refs.~\cite{DzuFlaSaf06,radpot}.

 The results for Cf$^{15+}$ and  Es$^{16+}$ ions are presented in
 Table \ref{t:Cf-en3}. 
Energy level diagrams for these ions are shown in Figs.~\ref{f:Cf15}
and  \ref{f:Es16}, respectively. 
Atomic properties of the clock transitions in Cf$^{15+}$,  Es$^{16+}$,
and  Es$^{17+}$ ions are listed in Table~\ref{t:CfEs-tran}. 
We also included Es$^{17+}$ ion properties in Table~\ref{t:CfEs-tran}
since it also satisfies the clock requirements, but has lower
sensitivity to $\alpha$-variation. 

\begin{table}
\caption{\label{t:Cf-en3}
Excitation energies (in cm$^{-1}$),
$g$-factors and lifetimes (in $s$) of the lowest states of Cf$^{15+}$
and Es$^{16+}$ ion. 
The numbers in brackets represent powers of 10.}
\begin{ruledtabular}
\begin{tabular}{rlc r ll}
\multicolumn{1}{c}{Ion}& \multicolumn{1}{c}{Conf.}&
\multicolumn{1}{c}{Term}&
\multicolumn{1}{c}{Energy}& \multicolumn{1}{c}{g-factors}&
\multicolumn{1}{c}{Lifetime}\\
\hline
  Cf$^{15+}$    &$5f6p^2$&$  ^2$F$^o_{5/2}$&       0 & 0.843 & \\
     &$5f^26p$&$  ^4$I$^o_{9/2}$&   12314 & 0.813 & 6.9[+3] \\
     &$5f6p^2$&$  ^2$F$^o_{7/2}$&   21947 & 1.083 & 1.2[-2]\\
     &$5f^26p$&$  ^2$F$^o_{5/2}$&   26665 & 0.715 & 0.26\\
     &$5f^26p$&$  ^2$D$^o_{3/2}$&  27750 & 0.765 & 1.9\\
     &$5f^26p$&$  ^2$G$^o_{7/2}$&   28875 & 0.868 & 1.6[-2]\\
     &$5f^26p$&$ ^4$I$^o_{11/2}$&   36564 & 0.996 & 3.1[-3]\\
     &$5f^26p$&$  ^4$H$^o_{9/2}$&   37392 & 1.029 & 1.5[-2] \\[0.5pc]
 Es$^{16+}$    & $5f^26p$       &$^4$I$^o_{9/2} $&     0&  0.808&  \\
  & $5f^26p$       &$^2$F$^o_{5/2} $&  6994&  0.788& 1.6[+4] \\
  & $5f^3  $       &$^2$H$^o_{9/2} $& 10591&  0.804& 3.4 \\
  & $5f^26p$     &$^2$D$^o_{3/2} $& 11056& 0.806&   666\\
  & $5f^26p$       &$^2$G$^o_{7/2} $& 15441&  0.842& 0.70 \\
  & $5f^26p$       &$^2$F$^o_{5/2} $& 22616&  0.778& 0.17 \\
  & $5f^26p$       &$^4$I$^o_{11/2}$& 24301&  0.990& 3.1[-3] \\
  & $5f^26p$       &$^4$H$^o_{9/2} $& 28351&  1.026& 1.2[-2] \\
\end{tabular}
\end{ruledtabular}
\end{table}

\begin{table*}
\caption{\label{t:CfEs-tran}
Clock transitions in Cf$^{15+}$, Es$^{16+}$ and Es$^{17+}$. $\alpha_0$
is scalar static polarisability (a.u.), $\lambda$ is the clock
transition wavelength, $\beta$ is relative frequency shift due to BBR,
$\tau$ is the lifetime of the clock state, $\omega$ is the frequency
of the clock transition, $q$ is the the coefficient of sensitivity to
$\alpha$-variation, $K$ is enhancement factor. 
The numbers in brackets represent powers of 10.}
\begin{ruledtabular}
\begin{tabular}{l llc llc rrccrr}
\multicolumn{1}{c}{Ion}&
\multicolumn{6}{c}{Clock transition}&
\multicolumn{1}{c}{$\tau$}&\multicolumn{1}{c}{$\lambda$}&
\multicolumn{1}{c}{$\hbar\omega$}&
\multicolumn{1}{c}{$\beta$}&
\multicolumn{1}{c}{$q$}&
\multicolumn{1}{c}{$K$}\\
&\multicolumn{2}{c}{Ground state}&\multicolumn{1}{c}{$\alpha_0$}&
\multicolumn{2}{c}{Clock state}&\multicolumn{1}{c}{$\alpha_0$}&
\multicolumn{1}{c}{s}&\multicolumn{1}{c}{nm}&\multicolumn{1}{c}{cm$^{-1}$}&
\multicolumn{1}{c}{}& \\
\hline
$^{249}$Cf$^{15+}$ & $5f6p^2$ & $ ^2$F$^o_{5/2}$ & 0.317 & $5f^26p$ & $
^4$I$^o_{9/2}$& 0.183 & 6900 &812&12314 & -2.9[-18] & 380000 & 57 \\

$^{253}$Es$^{16+}$ & $5f^26p$ & $ ^4$I$^o_{9/2}$&   0.161     & $5f^26p$ & $
^2$F$^o_{5/2}$& 0.199   & 16000 &1430& 6994 &   1.6[-18]   &   -184000 & -53 \\

$^{252}$Es$^{17+}$ & $5f^2$ & $^3$H$_{4}$& 0.032 & $5f6p$ & $^3$F$_{2}$ & 0.042 & 11000 &1343& 7445 &
4.0[-19] & -46600 & -13 \\

\end{tabular}
\end{ruledtabular}
\end{table*}
\begin{table}
\caption{\label{t:M1}
Strong magnetic dipole (M1) transitions involving ground state (GS) and clock state (CS) of Cf$^{15+}$, Es$^{16+}$ and Es$^{17+}$ ions. $A$ is the amplitude in Bohr magnetons $\mu$ ($\mu = e\hbar/2mc$), $\tau$ is the equivalent lifetime, which is the inverse of the rate of spontaneous emission from upper state to clock or ground state.}
\begin{ruledtabular}
\begin{tabular}{lllcrcr}
\multicolumn{1}{c}{Ion}&
\multicolumn{1}{c}{Lower}&
\multicolumn{1}{c}{Upper}&
\multicolumn{1}{c}{Wavelength}&
\multicolumn{1}{c}{Frequency}&
\multicolumn{1}{c}{$A$}&
\multicolumn{1}{c}{$\tau$}\\
&\multicolumn{1}{c}{State}&\multicolumn{1}{c}{State}&\multicolumn{1}{c}{nm}&
\multicolumn{1}{c}{(cm$^{-1}$)}&
\multicolumn{1}{c}{($\mu$)}&
\multicolumn{1}{c}{(ms)}\\
\hline
Cf$^{15+}$ & GS & $^2$F$^o_{7/2}$ &456& 21947 & 1.55 & 11 \\
                   & CS & $^4$I$^o_{11/2}$ &412& 24250 & 3.17 & 3 \\
Es$^{16+}$ & GS & $^4$I$^o_{11/2}$ &412& 24301 & 3.03 & 3 \\
                   & CS & $^2$F$^o_{7/2}$ &359& 27837 & 1.82 & 4 \\
Es$^{17+}$ & GS & $^3$H$_{5}$ &476& 21014 & 2.98 & 5 \\
\end{tabular}
\end{ruledtabular}
\end{table}

Clock transitions in Es$^{17+}$,  Cf$^{15+}$ and  Es$^{16+}$ ions are electric quadrupole transitions (E2) between ground and first excited state (see Figs.~\ref{f:Cf15} and \ref{f:Es16}). Probability of spontaneous emission from the clock to the ground state is given by (atomic units)
\begin{equation}
T^{\rm E2}_{c \to g} = \frac{1}{15}(\alpha\omega)^5\frac{\langle c||E2||g\rangle^2}{2J_c+1},
\label{eq:E2}
\end{equation}
where $\alpha$ is the fine structure constant, $\omega$ is the frequency of the transition, and $J_c$ is the total angular momentum of the clock state.
Corresponding lifetimes are presented in Table~\ref{t:CfEs-tran}. Note that they are either within or very close to the desirable range.

All considered here actinide ions have relatively strong magnetic dipole transitions to the ground and clock states. Note that since we are considering ions in the vicinity of the $6p-5f$ level crossing only states of the same parity lie within optical range. Therefore, the strongest optical transitions are always M1 transitions.  The rate of spontaneous emission from upper state $a$ to lower state $b$ is given by (atomic units)
\begin{equation}
T^{\rm M1}_{a \to b} = \frac{4}{3}(\alpha\omega)^3\frac{\langle a||M1||b\rangle^2}{2J_a+1}.
\label{eq:M1}
\end{equation}
The strongest M1 amplitudes are presented in Table~\ref{t:M1}.

Frequency shift due to black body radiation can be presented in a form
\begin{equation}
\delta\omega/\omega \approx \beta(T/T_0)^4,
\label{eq:BBR}
\end{equation}
where $T$ is temperature, $T_0$ is room temperature ($T_0$= 300 K), and parameter $\beta$ is given by
\begin{equation}
\beta = -\frac{\Delta \alpha_0}{2\omega}\left(831.9~ {\rm V/m}\right)^2.
\label{eq:beta}
\end{equation}
Here $\Delta \alpha_0$ is the difference in static scalar polarizabilities of clock and ground states, $\omega$ is the frequency of the clock transition.

Calculated polarizabilities and BBR shift parameter $\beta$ for considered here clock transitions are presented in
Table~\ref{t:CfEs-tran}. Note that the fractional BBR shift is of the the order of $10^{-18}$ even at 300~K. It can be further reduced by
 lowing the temperature.
\begin{table}
\caption{\label{t:CfEs-isotopes}
Long-living isotopes of Cf and Es with convenient values of nuclear
spin ($I$).}
\begin{ruledtabular}
\begin{tabular}{lcc}
\multicolumn{1}{c}{Isotope}&
\multicolumn{1}{c}{Lifetime}&
\multicolumn{1}{c}{$I$} \\
\hline
$^{249}$Cf & 351 y & 9/2 \\
$^{252}$Es & 1.29 y & 5 \\
$^{253}$Es & 20 d & 7/2 \\
$^{255}$Es & 40 d & 7/2 \\

\end{tabular}
\end{ruledtabular}
\end{table}
The clock states considered in this work have large value of the total
angular momentum $J$ (see Table~\ref{t:CfEs-tran}). As a result, they
have non-zero quadrupole moment that couples to gradient of external
electric fields. However, corresponding frequency shift can be avoided
by choosing appropriate components of the hyperfine structure of the
clock states. The quadrupole energy shift is proportional to $3M^2 -
F(F+1)$, where $F$ is total atomic angular momentum including nuclear
spin $I$ ($\mathbf{F}=\mathbf{J}+ \mathbf{I}$), $M$ is projection of
$F$ on $z$-axis. The shift is zero for states with $F=3$, $M=2$ or
$F=0$, $M=0$. It can also be numerically suppressed for appropriate
hyperfine structure
states~\cite{DDF-PRL12}. Table~\ref{t:CfEs-isotopes} lists isotopes of 
Cf and Es which have long lifetime and suitable values of nuclear
spin. Using the $^{249}$Cf  isotope for the Cf$^{15+}$ clock, the
$^{252}$Es  isotope for the Es$^{17+}$ clock and $^{253}$Es or
$^{255}$Es isotopes for the Es$^{16+}$ clock makes it always possible
to chose states with $F=3$, $M=2$ for both clock states. For upper
state of $^{249}$Cf$^{15+}$ clock the choice of $F=0$,  $M=0$ is also
possible. 

The results for $q$ and $K$ parameters of clock transitions are
presented in Table~\ref{t:CfEs-tran}. Note that $q$ values are very
large, with the different sign of $q$ for Cf and Es ions. The $6p$ -
$5f$ transition in the Cf$^{15+}$ ion becomes the $5f$ - $6p$
transition in the Es$^{17+}$ and Es$^{16+}$ ions, thus reversing the
sign of the $K$. While the leading configurations for both clock
states of the Es$^{16+}$ ion are the same the sensitivity coefficient
$q$ is still large. This results from  the strong admixture of the
$5f6p^2$ configuration, which is different for the lower and upper
states. If the ratio of the $^{253}$Es$^{16+}$ clock frequency and the
$^{249}$Cf$^{15+}$ clock frequency is measured, the sensitivity is given by
\begin{equation}
\frac{\partial}{\partial t}\ln\frac{\omega_2}{\omega_1} =
110\frac{1}{\alpha}\frac{\partial\alpha}{\partial t}.
\end{equation}
according to Eq.~\ref{e:ir17}.

In summary, we find that the pair of Cf$^{15+}$ and  Es$^{16+}$ ions
satisfy the  requirements for the design of a high-precision clock 
with the enhanced sensitivity to $\alpha$-variation of $\Delta
K=110$. Therefore, if $\alpha$ changes in time or with spatial
position, the ratio of the Cf$^{15+}$ and Es$^{16+}$ clock frequencies
changes more than one hundred times faster. This drastically reduces
the stringent requirement to the accuracy of the frequency
measurement needed to achieve sensitivity for the test of the spatial
$\alpha$-variation  indicated by the astrophysical observations
\cite{alpha-dipole}. We have surveyed a large number of potential
systems and large $K$ was either due to a small transition frequency
$\omega$ or occurred in systems that make design of accurate clock
difficult due to other reason, such a too short or too long clock
state lifetimes. To the best of our knowledge, this is the only pair
of ions with such a large $\delta K$ that also satisfy the clock
requirements introduced in \cite{Ho14+15}. 

We note that Cf$^{15+}$  is a particular good clock candidate with
optical (812~nm) clock transition, very long-lived isotope that
conveniently allows for $M=0$ $F=0$ selection of the clock state, and
very large $K=57$. Therefore, an experiment measuring  ratio of
Cf$^{15+}$ clock frequency with any other current clock where
10$^{-18}$ accuracy can be achieved will also provide excellent test
of $\alpha$-variation, which will lose only factor of two in
sensitivity to Cf/Es pair but would be easier to implement.

We thank Michiharu Wada and Hidetoshi Katori for stimulating discussions.
 M.S.S. thanks the School of Physics at UNSW, Sydney, Australia  for hospitality and
acknowledges support from the Gordon Godfrey Fellowship program, UNSW.
This work was supported by U.S. NSF grant No. PHY-1404156 and the Australian Research Council.

\bibliographystyle{apsrev}
\bibliography{dzuba,other}

\begin{thebibliography}{38}
\expandafter\ifx\csname natexlab\endcsname\relax\def\natexlab#1{#1}\fi
\expandafter\ifx\csname bibnamefont\endcsname\relax
  \def\bibnamefont#1{#1}\fi
\expandafter\ifx\csname bibfnamefont\endcsname\relax
  \def\bibfnamefont#1{#1}\fi
\expandafter\ifx\csname citenamefont\endcsname\relax
  \def\citenamefont#1{#1}\fi
\expandafter\ifx\csname url\endcsname\relax
  \def\url#1{\texttt{#1}}\fi
\expandafter\ifx\csname urlprefix\endcsname\relax\def\urlprefix{URL }\fi
\providecommand{\bibinfo}[2]{#2}
\providecommand{\eprint}[2][]{\url{#2}}

\bibitem[{\citenamefont{Uzan}(2003)}]{Uzan}
\bibinfo{author}{\bibfnamefont{J.-P.} \bibnamefont{Uzan}},
  \bibinfo{journal}{Rev. Mod. Phys.} \textbf{\bibinfo{volume}{75}},
  \bibinfo{pages}{403} (\bibinfo{year}{2003}).

\bibitem[{\citenamefont{{Uzan}}(2011)}]{Uzan11}
\bibinfo{author}{\bibfnamefont{J.-P.} \bibnamefont{{Uzan}}},
  \bibinfo{journal}{Living Reviews in Relativity}
  \textbf{\bibinfo{volume}{14}}, \bibinfo{pages}{2} (\bibinfo{year}{2011}).

\bibitem[{\citenamefont{Webb et~al.}(2011)\citenamefont{Webb, King, Murphy,
  Flambaum, Carswell, and Bainbridge}}]{alpha-dipole}
\bibinfo{author}{\bibfnamefont{J.~K.} \bibnamefont{Webb}},
  \bibinfo{author}{\bibfnamefont{J.~A.} \bibnamefont{King}},
  \bibinfo{author}{\bibfnamefont{M.~T.} \bibnamefont{Murphy}},
  \bibinfo{author}{\bibfnamefont{V.~V.} \bibnamefont{Flambaum}},
  \bibinfo{author}{\bibfnamefont{R.~F.} \bibnamefont{Carswell}},
  \bibnamefont{and} \bibinfo{author}{\bibfnamefont{M.~B.}
  \bibnamefont{Bainbridge}}, \bibinfo{journal}{Phys. Rev. Lett.}
  \textbf{\bibinfo{volume}{107}}, \bibinfo{pages}{191101}
  (\bibinfo{year}{2011}).

\bibitem[{\citenamefont{Berengut and Flambaum}(2012)}]{dipole-dot}
\bibinfo{author}{\bibfnamefont{J.}~\bibnamefont{Berengut}} \bibnamefont{and}
  \bibinfo{author}{\bibfnamefont{V.~V.} \bibnamefont{Flambaum}},
  \bibinfo{journal}{Europ. Phys. Lett.} \textbf{\bibinfo{volume}{97}},
  \bibinfo{pages}{20006} (\bibinfo{year}{2012}).

\bibitem[{\citenamefont{{Whitmore} and {Murphy}}(2015)}]{WhitMur15}
\bibinfo{author}{\bibfnamefont{J.~B.} \bibnamefont{{Whitmore}}}
  \bibnamefont{and} \bibinfo{author}{\bibfnamefont{M.~T.}
  \bibnamefont{{Murphy}}}, \bibinfo{journal}{Mon. Not. R. Astron. Soc.}
  \textbf{\bibinfo{volume}{447}}, \bibinfo{pages}{446} (\bibinfo{year}{2015}).

\bibitem[{\citenamefont{Leefer et~al.}(2013)\citenamefont{Leefer, Weber,
  Cing\"{o}z, Torgerson, and Budker}}]{Dy-alpha-dot}
\bibinfo{author}{\bibfnamefont{N.}~\bibnamefont{Leefer}},
  \bibinfo{author}{\bibfnamefont{C.~T.~M.} \bibnamefont{Weber}},
  \bibinfo{author}{\bibfnamefont{A.}~\bibnamefont{Cing\"{o}z}},
  \bibinfo{author}{\bibfnamefont{J.~R.} \bibnamefont{Torgerson}},
  \bibnamefont{and} \bibinfo{author}{\bibfnamefont{D.}~\bibnamefont{Budker}},
  \bibinfo{journal}{Phys. Rev. Lett.} \textbf{\bibinfo{volume}{111}},
  \bibinfo{pages}{060801} (\bibinfo{year}{2013}).

\bibitem[{\citenamefont{Rosenband et~al.}(2008)\citenamefont{Rosenband, Hume,
  Schmidt, Chou, Brush, Lorini, Oskay, Drullinger, Fortier, Stalnaker
  et~al.}}]{Al+}
\bibinfo{author}{\bibfnamefont{T.}~\bibnamefont{Rosenband}},
  \bibinfo{author}{\bibfnamefont{D.~B.} \bibnamefont{Hume}},
  \bibinfo{author}{\bibfnamefont{P.~O.} \bibnamefont{Schmidt}},
  \bibinfo{author}{\bibfnamefont{C.~W.} \bibnamefont{Chou}},
  \bibinfo{author}{\bibfnamefont{A.}~\bibnamefont{Brush}},
  \bibinfo{author}{\bibfnamefont{L.}~\bibnamefont{Lorini}},
  \bibinfo{author}{\bibfnamefont{W.~H.} \bibnamefont{Oskay}},
  \bibinfo{author}{\bibfnamefont{R.~E.} \bibnamefont{Drullinger}},
  \bibinfo{author}{\bibfnamefont{T.~M.} \bibnamefont{Fortier}},
  \bibinfo{author}{\bibfnamefont{J.~E.} \bibnamefont{Stalnaker}},
  \bibnamefont{et~al.}, \bibinfo{journal}{Science}
  \textbf{\bibinfo{volume}{319}}, \bibinfo{pages}{1808} (\bibinfo{year}{2008}).

\bibitem[{\citenamefont{Godun et~al.}(2014)\citenamefont{Godun, Nisbet-Jones,
  Jones, King, Johnson, Margolis, Szymaniec, Lea, Bongs, and
  Gill}}]{Gill-Yb+14}
\bibinfo{author}{\bibfnamefont{R.}~\bibnamefont{Godun}},
  \bibinfo{author}{\bibfnamefont{P.~B.~R.} \bibnamefont{Nisbet-Jones}},
  \bibinfo{author}{\bibfnamefont{J.~M.} \bibnamefont{Jones}},
  \bibinfo{author}{\bibfnamefont{S.~A.} \bibnamefont{King}},
  \bibinfo{author}{\bibfnamefont{L.~A.~M.} \bibnamefont{Johnson}},
  \bibinfo{author}{\bibfnamefont{H.~S.} \bibnamefont{Margolis}},
  \bibinfo{author}{\bibfnamefont{K.}~\bibnamefont{Szymaniec}},
  \bibinfo{author}{\bibfnamefont{S.~N.} \bibnamefont{Lea}},
  \bibinfo{author}{\bibfnamefont{K.}~\bibnamefont{Bongs}}, \bibnamefont{and}
  \bibinfo{author}{\bibfnamefont{P.}~\bibnamefont{Gill}},
  \bibinfo{journal}{Phys. Rev. Lett.} \textbf{\bibinfo{volume}{113}},
  \bibinfo{pages}{210801} (\bibinfo{year}{2014}).

\bibitem[{\citenamefont{Huntemann et~al.}(2014)\citenamefont{Huntemann,
  Lipphardt, Tamm, Gerginov, Weyers, and Peik}}]{Peik-Yb+14}
\bibinfo{author}{\bibfnamefont{N.}~\bibnamefont{Huntemann}},
  \bibinfo{author}{\bibfnamefont{B.}~\bibnamefont{Lipphardt}},
  \bibinfo{author}{\bibfnamefont{C.}~\bibnamefont{Tamm}},
  \bibinfo{author}{\bibfnamefont{V.}~\bibnamefont{Gerginov}},
  \bibinfo{author}{\bibfnamefont{S.}~\bibnamefont{Weyers}}, \bibnamefont{and}
  \bibinfo{author}{\bibfnamefont{E.}~\bibnamefont{Peik}},
  \bibinfo{journal}{Phys. Rev. Lett.} \textbf{\bibinfo{volume}{113}},
  \bibinfo{pages}{210802} (\bibinfo{year}{2014}).

\bibitem[{\citenamefont{{Nicholson} et~al.}(2015)\citenamefont{{Nicholson},
  {Campbell}, {Hutson}, {Marti}, {Bloom}, {McNally}, {Zhang}, {Barrett},
  {Safronova}, {Strouse} et~al.}}]{NicCamHut15}
\bibinfo{author}{\bibfnamefont{T.~L.} \bibnamefont{{Nicholson}}},
  \bibinfo{author}{\bibfnamefont{S.~L.} \bibnamefont{{Campbell}}},
  \bibinfo{author}{\bibfnamefont{R.~B.} \bibnamefont{{Hutson}}},
  \bibinfo{author}{\bibfnamefont{G.~E.} \bibnamefont{{Marti}}},
  \bibinfo{author}{\bibfnamefont{B.~J.} \bibnamefont{{Bloom}}},
  \bibinfo{author}{\bibfnamefont{R.~L.} \bibnamefont{{McNally}}},
  \bibinfo{author}{\bibfnamefont{W.}~\bibnamefont{{Zhang}}},
  \bibinfo{author}{\bibfnamefont{M.~D.} \bibnamefont{{Barrett}}},
  \bibinfo{author}{\bibfnamefont{M.~S.} \bibnamefont{{Safronova}}},
  \bibinfo{author}{\bibfnamefont{G.~F.} \bibnamefont{{Strouse}}},
  \bibnamefont{et~al.}, \bibinfo{journal}{Nature Commun.}
  \textbf{\bibinfo{volume}{6}}, \bibinfo{eid}{6896} (\bibinfo{year}{2015}).

\bibitem[{\citenamefont{Bloom et~al.}(2014)\citenamefont{Bloom, Nicholson,
  Campbell, Bishof, Zhang, Zhang, Bromley, and Ye}}]{Sr-clock14}
\bibinfo{author}{\bibfnamefont{B.~J.} \bibnamefont{Bloom}},
  \bibinfo{author}{\bibfnamefont{T.~L.} \bibnamefont{Nicholson}},
  \bibinfo{author}{\bibfnamefont{J.~R. W. S.~L.} \bibnamefont{Campbell}},
  \bibinfo{author}{\bibfnamefont{M.}~\bibnamefont{Bishof}},
  \bibinfo{author}{\bibfnamefont{X.}~\bibnamefont{Zhang}},
  \bibinfo{author}{\bibfnamefont{W.}~\bibnamefont{Zhang}},
  \bibinfo{author}{\bibfnamefont{S.~L.} \bibnamefont{Bromley}},
  \bibnamefont{and} \bibinfo{author}{\bibfnamefont{J.}~\bibnamefont{Ye}},
  \bibinfo{journal}{Nature} \textbf{\bibinfo{volume}{506}}, \bibinfo{pages}{71}
  (\bibinfo{year}{2014}).

\bibitem[{\citenamefont{Ushijima et~al.}(2015)\citenamefont{Ushijima, Takamoto,
  Das, Ohkubo, and Katori}}]{Katori}
\bibinfo{author}{\bibfnamefont{I.}~\bibnamefont{Ushijima}},
  \bibinfo{author}{\bibfnamefont{M.}~\bibnamefont{Takamoto}},
  \bibinfo{author}{\bibfnamefont{M.}~\bibnamefont{Das}},
  \bibinfo{author}{\bibfnamefont{T.}~\bibnamefont{Ohkubo}}, \bibnamefont{and}
  \bibinfo{author}{\bibfnamefont{H.}~\bibnamefont{Katori}},
  \bibinfo{journal}{Nature Photonics} \textbf{\bibinfo{volume}{9}},
  \bibinfo{pages}{185} (\bibinfo{year}{2015}).

\bibitem[{\citenamefont{Angstmann et~al.}(2004)\citenamefont{Angstmann, Dzuba,
  and Flambaum}}]{AngDzuFla04}
\bibinfo{author}{\bibfnamefont{E.~J.} \bibnamefont{Angstmann}},
  \bibinfo{author}{\bibfnamefont{V.~A.} \bibnamefont{Dzuba}}, \bibnamefont{and}
  \bibinfo{author}{\bibfnamefont{V.~V.} \bibnamefont{Flambaum}},
  \bibinfo{journal}{Phys. Rev. A} \textbf{\bibinfo{volume}{70}},
  \bibinfo{pages}{014102} (\bibinfo{year}{2004}).

\bibitem[{\citenamefont{Dzuba et~al.}(1999{\natexlab{a}})\citenamefont{Dzuba,
  Flambaum, and Webb}}]{DzuFlaWeb99}
\bibinfo{author}{\bibfnamefont{V.~A.} \bibnamefont{Dzuba}},
  \bibinfo{author}{\bibfnamefont{V.~V.} \bibnamefont{Flambaum}},
  \bibnamefont{and} \bibinfo{author}{\bibfnamefont{J.~K.} \bibnamefont{Webb}},
  \bibinfo{journal}{Phys. Rev. Lett.} \textbf{\bibinfo{volume}{82}},
  \bibinfo{pages}{888} (\bibinfo{year}{1999}{\natexlab{a}}).

\bibitem[{\citenamefont{Dzuba et~al.}(1999{\natexlab{b}})\citenamefont{Dzuba,
  Flambaum, and Webb}}]{DzuFlaWeb99a}
\bibinfo{author}{\bibfnamefont{V.~A.} \bibnamefont{Dzuba}},
  \bibinfo{author}{\bibfnamefont{V.~V.} \bibnamefont{Flambaum}},
  \bibnamefont{and} \bibinfo{author}{\bibfnamefont{J.~K.} \bibnamefont{Webb}},
  \bibinfo{journal}{Phys. Rev. A} \textbf{\bibinfo{volume}{59}},
  \bibinfo{pages}{230} (\bibinfo{year}{1999}{\natexlab{b}}).

\bibitem[{\citenamefont{Flambaum and Dzuba}(2009)}]{FlaDzuCJP}
\bibinfo{author}{\bibfnamefont{V.~V.} \bibnamefont{Flambaum}} \bibnamefont{and}
  \bibinfo{author}{\bibfnamefont{V.~A.} \bibnamefont{Dzuba}},
  \bibinfo{journal}{Can. J. Phys.} \textbf{\bibinfo{volume}{87}},
  \bibinfo{pages}{25} (\bibinfo{year}{2009}).

\bibitem[{\citenamefont{{Ludlow} et~al.}(2015)\citenamefont{{Ludlow}, {Boyd},
  {Ye}, {Peik}, and {Schmidt}}}]{LudBoyYe15}
\bibinfo{author}{\bibfnamefont{A.~D.} \bibnamefont{{Ludlow}}},
  \bibinfo{author}{\bibfnamefont{M.~M.} \bibnamefont{{Boyd}}},
  \bibinfo{author}{\bibfnamefont{J.}~\bibnamefont{{Ye}}},
  \bibinfo{author}{\bibfnamefont{E.}~\bibnamefont{{Peik}}}, \bibnamefont{and}
  \bibinfo{author}{\bibfnamefont{P.~O.} \bibnamefont{{Schmidt}}},
  \bibinfo{journal}{Rev. Mod. Phys.} \textbf{\bibinfo{volume}{87}},
  \bibinfo{pages}{637} (\bibinfo{year}{2015}).

\bibitem[{\citenamefont{Safronova
  et~al.}(2014{\natexlab{a}})\citenamefont{Safronova, Dzuba, Flambaum,
  Safronova, Porsev, and Kozlov}}]{HCI-PRL14}
\bibinfo{author}{\bibfnamefont{M.~S.} \bibnamefont{Safronova}},
  \bibinfo{author}{\bibfnamefont{V.~A.} \bibnamefont{Dzuba}},
  \bibinfo{author}{\bibfnamefont{V.~V.} \bibnamefont{Flambaum}},
  \bibinfo{author}{\bibfnamefont{U.~I.} \bibnamefont{Safronova}},
  \bibinfo{author}{\bibfnamefont{S.~G.} \bibnamefont{Porsev}},
  \bibnamefont{and} \bibinfo{author}{\bibfnamefont{M.~G.}
  \bibnamefont{Kozlov}}, \bibinfo{journal}{Phys. Rev. Lett.}
  \textbf{\bibinfo{volume}{113}}, \bibinfo{pages}{030801}
  (\bibinfo{year}{2014}{\natexlab{a}}).

\bibitem[{\citenamefont{Berengut et~al.}(2012)\citenamefont{Berengut, Dzuba,
  Flambaum, and Ong}}]{BDFO-Cf12}
\bibinfo{author}{\bibfnamefont{J.~C.} \bibnamefont{Berengut}},
  \bibinfo{author}{\bibfnamefont{V.~A.} \bibnamefont{Dzuba}},
  \bibinfo{author}{\bibfnamefont{V.~V.} \bibnamefont{Flambaum}},
  \bibnamefont{and} \bibinfo{author}{\bibfnamefont{A.}~\bibnamefont{Ong}},
  \bibinfo{journal}{Phys. Rev. Lett.} \textbf{\bibinfo{volume}{109}},
  \bibinfo{pages}{070802} (\bibinfo{year}{2012}).

\bibitem[{\citenamefont{Berengut et~al.}(2010)\citenamefont{Berengut, Dzuba,
  and Flambaum}}]{BDF-HCI10}
\bibinfo{author}{\bibfnamefont{J.~C.} \bibnamefont{Berengut}},
  \bibinfo{author}{\bibfnamefont{V.~A.} \bibnamefont{Dzuba}}, \bibnamefont{and}
  \bibinfo{author}{\bibfnamefont{V.~V.} \bibnamefont{Flambaum}},
  \bibinfo{journal}{Phys. Rev. Lett.} \textbf{\bibinfo{volume}{105}},
  \bibinfo{pages}{120801} (\bibinfo{year}{2010}).

\bibitem[{\citenamefont{Dzuba et~al.}(2015)\citenamefont{Dzuba, Flambaum, and
  Katori}}]{Ho14+15}
\bibinfo{author}{\bibfnamefont{V.~A.} \bibnamefont{Dzuba}},
  \bibinfo{author}{\bibfnamefont{V.~V.} \bibnamefont{Flambaum}},
  \bibnamefont{and} \bibinfo{author}{\bibfnamefont{H.}~\bibnamefont{Katori}},
  \bibinfo{journal}{Phys. Rev. A} \textbf{\bibinfo{volume}{91}},
  \bibinfo{pages}{022119} (\bibinfo{year}{2015}).

\bibitem[{\citenamefont{Flambaum}(2006)}]{Fla-Th229-05}
\bibinfo{author}{\bibfnamefont{V.~V.} \bibnamefont{Flambaum}},
  \bibinfo{journal}{Phys. Rev. Lett.} \textbf{\bibinfo{volume}{97}},
  \bibinfo{pages}{092502} (\bibinfo{year}{2006}).

\bibitem[{\citenamefont{Berengut et~al.}(2011)\citenamefont{Berengut, Dzuba,
  Flambaum, and Ong}}]{BDFO-hole11}
\bibinfo{author}{\bibfnamefont{J.~C.} \bibnamefont{Berengut}},
  \bibinfo{author}{\bibfnamefont{V.~A.} \bibnamefont{Dzuba}},
  \bibinfo{author}{\bibfnamefont{V.~V.} \bibnamefont{Flambaum}},
  \bibnamefont{and} \bibinfo{author}{\bibfnamefont{A.}~\bibnamefont{Ong}},
  \bibinfo{journal}{Phys. Rev. Lett.} \textbf{\bibinfo{volume}{106}},
  \bibinfo{pages}{210802} (\bibinfo{year}{2011}).

\bibitem[{\citenamefont{Dzuba et~al.}(2012)\citenamefont{Dzuba, Derevianko, and
  Flambaum}}]{DDF12}
\bibinfo{author}{\bibfnamefont{V.~A.} \bibnamefont{Dzuba}},
  \bibinfo{author}{\bibfnamefont{A.}~\bibnamefont{Derevianko}},
  \bibnamefont{and} \bibinfo{author}{\bibfnamefont{V.~V.}
  \bibnamefont{Flambaum}}, \bibinfo{journal}{Phys. Rev. A}
  \textbf{\bibinfo{volume}{86}}, \bibinfo{pages}{054502}
  (\bibinfo{year}{2012}).

\bibitem[{\citenamefont{Safronova
  et~al.}(2014{\natexlab{b}})\citenamefont{Safronova, Dzuba, Flambaum,
  Safronova, Porsev, and Kozlov}}]{Ag-likeHCI14}
\bibinfo{author}{\bibfnamefont{M.~S.} \bibnamefont{Safronova}},
  \bibinfo{author}{\bibfnamefont{V.~A.} \bibnamefont{Dzuba}},
  \bibinfo{author}{\bibfnamefont{V.~V.} \bibnamefont{Flambaum}},
  \bibinfo{author}{\bibfnamefont{U.~I.} \bibnamefont{Safronova}},
  \bibinfo{author}{\bibfnamefont{S.~G.} \bibnamefont{Porsev}},
  \bibnamefont{and} \bibinfo{author}{\bibfnamefont{M.~G.}
  \bibnamefont{Kozlov}}, \bibinfo{journal}{Phys. Rev. A}
  \textbf{\bibinfo{volume}{90}}, \bibinfo{pages}{042513}
  (\bibinfo{year}{2014}{\natexlab{b}}).

\bibitem[{\citenamefont{Safronova
  et~al.}(2014{\natexlab{c}})\citenamefont{Safronova, Dzuba, Flambaum,
  Safronova, Porsev, and Kozlov}}]{Cd-likeHCI14}
\bibinfo{author}{\bibfnamefont{M.~S.} \bibnamefont{Safronova}},
  \bibinfo{author}{\bibfnamefont{V.~A.} \bibnamefont{Dzuba}},
  \bibinfo{author}{\bibfnamefont{V.~V.} \bibnamefont{Flambaum}},
  \bibinfo{author}{\bibfnamefont{U.~I.} \bibnamefont{Safronova}},
  \bibinfo{author}{\bibfnamefont{S.~G.} \bibnamefont{Porsev}},
  \bibnamefont{and} \bibinfo{author}{\bibfnamefont{M.~G.}
  \bibnamefont{Kozlov}}, \bibinfo{journal}{Phys. Rev. A}
  \textbf{\bibinfo{volume}{90}}, \bibinfo{pages}{052509}
  (\bibinfo{year}{2014}{\natexlab{c}}).

\bibitem[{\citenamefont{Windberger et~al.}(2015)\citenamefont{Windberger,
  L—pez-Urrutia, Bekker, Oreshkina, Berengut, Bock, Borschevsky, Dzuba, Eliav,
  Harman et~al.}}]{IrPRL15}
\bibinfo{author}{\bibfnamefont{A.}~\bibnamefont{Windberger}},
  \bibinfo{author}{\bibfnamefont{J.~R.~C.} \bibnamefont{L—pez-Urrutia}},
  \bibinfo{author}{\bibfnamefont{H.}~\bibnamefont{Bekker}},
  \bibinfo{author}{\bibfnamefont{N.}~\bibnamefont{Oreshkina}},
  \bibinfo{author}{\bibfnamefont{J.}~\bibnamefont{Berengut}},
  \bibinfo{author}{\bibfnamefont{V.}~\bibnamefont{Bock}},
  \bibinfo{author}{\bibfnamefont{A.}~\bibnamefont{Borschevsky}},
  \bibinfo{author}{\bibfnamefont{V.}~\bibnamefont{Dzuba}},
  \bibinfo{author}{\bibfnamefont{E.}~\bibnamefont{Eliav}},
  \bibinfo{author}{\bibfnamefont{Z.}~\bibnamefont{Harman}},
  \bibnamefont{et~al.}, \bibinfo{journal}{Phys. Rev. Lett.}
  \textbf{\bibinfo{volume}{114}}, \bibinfo{pages}{150801}
  (\bibinfo{year}{2015}).

\bibitem[{\citenamefont{Schm\"{o}ger et~al.}(2015)\citenamefont{Schm\"{o}ger,
  Versolato, Schwarz, Kohnen, Windberger, Piest, Feuchtenbeiner,
  Pedregosa-Gutierrez, Leopold, Micke et~al.}}]{SchVerSch15}
\bibinfo{author}{\bibfnamefont{L.}~\bibnamefont{Schm\"{o}ger}},
  \bibinfo{author}{\bibfnamefont{O.~O.} \bibnamefont{Versolato}},
  \bibinfo{author}{\bibfnamefont{M.}~\bibnamefont{Schwarz}},
  \bibinfo{author}{\bibfnamefont{M.}~\bibnamefont{Kohnen}},
  \bibinfo{author}{\bibfnamefont{A.}~\bibnamefont{Windberger}},
  \bibinfo{author}{\bibfnamefont{B.}~\bibnamefont{Piest}},
  \bibinfo{author}{\bibfnamefont{S.}~\bibnamefont{Feuchtenbeiner}},
  \bibinfo{author}{\bibfnamefont{J.}~\bibnamefont{Pedregosa-Gutierrez}},
  \bibinfo{author}{\bibfnamefont{T.}~\bibnamefont{Leopold}},
  \bibinfo{author}{\bibfnamefont{P.}~\bibnamefont{Micke}},
  \bibnamefont{et~al.}, \bibinfo{journal}{Science}
  \textbf{\bibinfo{volume}{347}}, \bibinfo{pages}{1233} (\bibinfo{year}{2015}).

\bibitem[{\citenamefont{Runke et~al.}(2014)\citenamefont{Runke, Dullmann,
  Eberhardt, Ellison, Gregorich, Hofmann, Jager, Kindler, Kratz, Krier
  et~al.}}]{Cf-prod}
\bibinfo{author}{\bibfnamefont{J.}~\bibnamefont{Runke}},
  \bibinfo{author}{\bibfnamefont{C.~E.} \bibnamefont{Dullmann}},
  \bibinfo{author}{\bibfnamefont{K.}~\bibnamefont{Eberhardt}},
  \bibinfo{author}{\bibfnamefont{P.~A.} \bibnamefont{Ellison}},
  \bibinfo{author}{\bibfnamefont{K.~E.} \bibnamefont{Gregorich}},
  \bibinfo{author}{\bibfnamefont{S.}~\bibnamefont{Hofmann}},
  \bibinfo{author}{\bibfnamefont{E.}~\bibnamefont{Jager}},
  \bibinfo{author}{\bibfnamefont{B.}~\bibnamefont{Kindler}},
  \bibinfo{author}{\bibfnamefont{J.~V.} \bibnamefont{Kratz}},
  \bibinfo{author}{\bibfnamefont{J.}~\bibnamefont{Krier}},
  \bibnamefont{et~al.}, \bibinfo{journal}{J. Radioanal. Nucl. Chem.}
  \textbf{\bibinfo{volume}{299}}, \bibinfo{pages}{1081} (\bibinfo{year}{2014}).

\bibitem[{\citenamefont{Meierfrankenfeld
  et~al.}(2011)\citenamefont{Meierfrankenfeld, Bury, and
  Thoennessen}}]{Es-prod}
\bibinfo{author}{\bibfnamefont{D.}~\bibnamefont{Meierfrankenfeld}},
  \bibinfo{author}{\bibfnamefont{A.}~\bibnamefont{Bury}}, \bibnamefont{and}
  \bibinfo{author}{\bibfnamefont{M.}~\bibnamefont{Thoennessen}},
  \bibinfo{journal}{Atomic Data and Nuclear Data Tables}
  \textbf{\bibinfo{volume}{97}}, \bibinfo{pages}{134} (\bibinfo{year}{2011}).

\bibitem[{\citenamefont{Safronova et~al.}(2009)\citenamefont{Safronova, Kozlov,
  Johnson, and Jiang}}]{SD+CI}
\bibinfo{author}{\bibfnamefont{M.~S.} \bibnamefont{Safronova}},
  \bibinfo{author}{\bibfnamefont{M.~G.} \bibnamefont{Kozlov}},
  \bibinfo{author}{\bibfnamefont{W.~R.} \bibnamefont{Johnson}},
  \bibnamefont{and} \bibinfo{author}{\bibfnamefont{D.}~\bibnamefont{Jiang}},
  \bibinfo{journal}{Phys. Rev. A} \textbf{\bibinfo{volume}{80}},
  \bibinfo{pages}{012516} (\bibinfo{year}{2009}).

\bibitem[{\citenamefont{Dzuba}(2014)}]{Dzu-CI-SD14}
\bibinfo{author}{\bibfnamefont{V.~A.} \bibnamefont{Dzuba}},
  \bibinfo{journal}{Phys. Rev. A} \textbf{\bibinfo{volume}{90}},
  \bibinfo{pages}{012517} (\bibinfo{year}{2014}).

\bibitem[{\citenamefont{Blundell et~al.}(1989)\citenamefont{Blundell, Johnson,
  Liu, and Sapirstein}}]{CC1}
\bibinfo{author}{\bibfnamefont{S.~A.} \bibnamefont{Blundell}},
  \bibinfo{author}{\bibfnamefont{W.~R.} \bibnamefont{Johnson}},
  \bibinfo{author}{\bibfnamefont{Z.~W.} \bibnamefont{Liu}}, \bibnamefont{and}
  \bibinfo{author}{\bibfnamefont{J.}~\bibnamefont{Sapirstein}},
  \bibinfo{journal}{Phys. Rev. A} \textbf{\bibinfo{volume}{40}},
  \bibinfo{pages}{2233} (\bibinfo{year}{1989}).

\bibitem[{\citenamefont{Johnson and Sapirstein}(1986)}]{B-splines}
\bibinfo{author}{\bibfnamefont{W.~R.} \bibnamefont{Johnson}} \bibnamefont{and}
  \bibinfo{author}{\bibfnamefont{J.}~\bibnamefont{Sapirstein}},
  \bibinfo{journal}{Phys. Rev. Lett.} \textbf{\bibinfo{volume}{57}},
  \bibinfo{pages}{1126} (\bibinfo{year}{1986}).

\bibitem[{\citenamefont{Dzuba}(2005)}]{Dzu05}
\bibinfo{author}{\bibfnamefont{V.~A.} \bibnamefont{Dzuba}},
  \bibinfo{journal}{Phys. Rev. A} \textbf{\bibinfo{volume}{71}},
  \bibinfo{pages}{032512} (\bibinfo{year}{2005}).

\bibitem[{\citenamefont{Dzuba et~al.}(2006)\citenamefont{Dzuba, Flambaum, and
  Safronova}}]{DzuFlaSaf06}
\bibinfo{author}{\bibfnamefont{V.~A.} \bibnamefont{Dzuba}},
  \bibinfo{author}{\bibfnamefont{V.~V.} \bibnamefont{Flambaum}},
  \bibnamefont{and} \bibinfo{author}{\bibfnamefont{M.~S.}
  \bibnamefont{Safronova}}, \bibinfo{journal}{Phys. Rev. A}
  \textbf{\bibinfo{volume}{73}}, \bibinfo{pages}{022112}
  (\bibinfo{year}{2006}).

\bibitem[{\citenamefont{Flambaum and Ginges}(2005)}]{radpot}
\bibinfo{author}{\bibfnamefont{V.~V.} \bibnamefont{Flambaum}} \bibnamefont{and}
  \bibinfo{author}{\bibfnamefont{J.~S.~M.} \bibnamefont{Ginges}},
  \bibinfo{journal}{Phys. Rev. A} \textbf{\bibinfo{volume}{72}},
  \bibinfo{pages}{052115} (\bibinfo{year}{2005}).

\bibitem[{\citenamefont{Derevianko et~al.}(2012)\citenamefont{Derevianko,
  Dzuba, and Flambaum}}]{DDF-PRL12}
\bibinfo{author}{\bibfnamefont{A.}~\bibnamefont{Derevianko}},
  \bibinfo{author}{\bibfnamefont{V.~A.} \bibnamefont{Dzuba}}, \bibnamefont{and}
  \bibinfo{author}{\bibfnamefont{V.~V.} \bibnamefont{Flambaum}},
  \bibinfo{journal}{Phys. Rev. Lett.} \textbf{\bibinfo{volume}{109}},
  \bibinfo{pages}{180801} (\bibinfo{year}{2012}).

\end{thebibliography}

\end{document}